# VIDEO SURVEILLANCE IN THE CLOUD?

DJ Neal[1] and Syed (Shawon) Rahman, Ph.D.[2]

[1]Information Assurance and Security, Capella University, Minneapolis, MN, USA
`dj@neal.ws`
[2]Assistant Professor, University of Hawaii-Hilo, HI USA and Adjunct Faculty, Capella University, Minneapolis, MN, USA
`syed.rahman@capella.edu`

## ABSTRACT

*A high-resolution video surveillance management system incurs huge amounts of storage and network bandwidth. The current infrastructure required to support a high-resolution video surveillance management system (VMS) is expensive and time consuming to plan, implement and maintain. With the recent advances in cloud technologies, opportunity for the utilization of virtualization and the opportunity for distributed computing techniques of cloud storage have been pursued on the basis to find out if the various cloud computing services that are available can support the current requirements to a high-resolution video surveillance management system. The research concludes, after investigating and comparing various Software as a Service (SaaS), Platform as a Service (PaaS), and Infrastructure as a Service (IaaS) cloud computing provides what is possible to architect a VMS using cloud technologies; however, it is more expensive and it will require additional reviews for legal implications, as well as emerging threats and countermeasures associated with using cloud technologies for a video surveillance management system..*

## 1. INTRODUCTION

In today's enterprise environment, surveillance systems contain a large number of cameras. Video surveillance systems have grown from the original closed-circuit video transmission (CCTV) environments into the self-contained digital video recorder (DVR) environments and now into the centrally managed Internet Protocol (IP) cameras, which can send the video anywhere that is on the internet including mobile devices and phones. Generally with IP cameras, the cameras point back to a centralized video management system (VMS), which is used to view, playback, and record the video. Research illustrates just how much data is required to support a high-resolution video surveillance management system and how it relates directly to using various cloud- computing provider for the possibility to plan, to deploy and to maintain such a high resource application system. Even though it is possible to use the current cloud computing resources of today, it is notsometimes economically sound to do so with an application that has such a high demand for storage and bandwidth.

Enterprises around the world spend a lot of money and resources on a video surveillance system, which includes the backend network system, storage system, and the computing infrastructure system, thus supporting the whole system as a whole. As cloud technologies gain more popularity, at what point does it become reasonable to use cloud technologies. An assessment was conducted to determine if it is practical for a video surveillance system to use various cloud technologies and answer the following questions:

- Can we put a video surveillance system on the cloud?
- Can the cloud save money?
- By using the new cloud technologies, does it remove some risks just to gain different risks later?

## 2. ASSUMPTIONS AND LIMITATIONS

Managing a VMS that utilizes cloud technologies is going to have an inherited risk of availability with the assumption that network connectivity is going to be required at all times. This could potentially place the VMS system vulnerable to Denial of Service attacks (DoS), which would lead to another layer of defense to handle such events. Additionally, with any VMS, issues could arise for managing the people and technology resources that are used to support the VMS. By using cloud technologies within a VMS, it is going to require another layer of management for controlling the various access requirements as users and administrators within the VMS. Therefore, information security control mechanisms such as physical, technical and administrative, needs to be documented and implemented in ways to prevent and detect the correct forms of access controls and to be flexible enough to integrate them into the company's culture. Due to limited availability of wireless cameras, only Ethernet cameras are going to be used. Internet Service Provider costs and fees are not being included in any calculations and that each location within the enterprise is going to have internet service available. Finally, there is the assumption that all cameras have been purchased in the past and so the cost for cameras will not be included in any calculations unless otherwise indicated.

## 3. ETHICAL AND LEGAL IMPLICATIONS

There can be various ethical and legal implications toward storing video surveillance footage off site and into a cloud infrastructure. Cloud providers are going to be required to be compliant and certified in regards to various laws and practices such as the Health Insurance Portability and Accountability Act (HIPPA) and the Payment Card Industry Data Security Standard (PCI DSS). There can be risk with Personal Privacy Information if video surveillance footage is not properly protected by following sound workflows and implementing common sense which could alter any infrastructure. Additionally, one main purpose of any VMS is to extract video footage to be used in a court of law. There could be legal implication having a VMS system that is using cloud technologies that cannot properly impound evidence or preserve the chain of custody for any video evidence. Courts have to address numerous legal issues when dealing with video as evidence, especially now with the effectiveness of ease of video surveillance systems that are easily concealable, and with advances by new technologies including high-resolution [1]. Most states, including Florida, have Rules of Civil Procedures, which are designed to provide protection from the use of "surprise, trickery, bluff and legal gymnastics when using video as evidence"[1]. Therefore, it is important to establish early on if video surveillance that is collected as evidence is considered as non-work product or work product based on the investigation. Evidence can have a significant different value in a court case depending on if it is considered a non-work product or a work product. Non-work products include video footage from a static or permanent video surveillance system which is installed on the premises, rather than a work product which includes video footage from a private investigator or another source that is not necessarily there to be a surveillance system [1]. Another challenge in the courts for surveillance video evidence is that it's not considered hearsay and therefore has to be authenticated similar to photographs [1]. Authentication of video can be done by having the videographer testify what is exactly in the video, a person that is in the video confirms that they are the ones in the video, or have a witness that is in the video confirms the individual that are in the video. Additionally, if no witness is able to authenticate the surveillance video than under the "silent witness" theory a judge can determine if the video can be authenticated if the following requirements are met [1]:

- there is evidence establishing the time and date of the video;
- there was no tampering with the video;
- the video equipment used was sound; and
- there is testimony identifying the participants depicted in the video.

It is also possible that a judge could determine that the surveillance video adds more confusion to the case and might just request a still-frame photo extracted from the video surveillance to prevent the danger of unfair prejudice or misleading the jury[1]. Therefore, it is important that any video management system using cloud technologies be able to export video surveillance footage or a still-frame photo from the same video surveillance footage.

## 4. RISK ASSESSMENT

As a fast-growing technology, cloud technologies in the industry do not always adhere to standard matrix, terminology, and services. As with any man-made devices technology, computing systems are acceptable to failure and security compromise. It is a paradigm for cloud computing services to provide large-scale computations, dynamic configurations, measured services and on-demand self-service adaptation that hides the inner workings and complexities from the subscriber[2]. However, when there are workings and complexities that are hidden, a level of trust has to be established to ensure that the risks associated with using cloud technologies is at a minimum. Exactly how does one verify that all risk are at a minimum when using cloud technologies? That is the open question for subscribers that use cloud technologies. As with any moving target, a risk assessment today will not necessarily be the same risk assessment tomorrow. So for a baseline risk assessment, using cloud computing technologies there are the following categories, which comes from the NIST Special Publication 146 will be used [3]:

- Computing Performance
- Cloud Reliability
- Economic Goals
- Compliance
- Data and Application Security

## 5. COMPUTING PERFORMANCE

Computing performance when being performed within a remote location incurs the same performance issues and standards when performing computations locally. However, with cloud computing there is inherently at least one Internet round-trip time lapse that is not necessarily under the control of a provider or subscriber and which can place any cloud application at a higher risk by having variations in network service response times [3]. Additionally, to leverage the general cloud computing usage to provide task parallelism and rapid elasticity growth is going to require a higher quality toward scalable programming in order to fully take advantage of the on-demand cloud computing architectures [3]. Finally, with computing performance there is of course managing the data storage, requesting additional capacity, physical location restrictions, verification that data is deleted securely, and managing and administering access control privileges with external parties [3].

## 6. CLOUD RELIABILITY

Any system cloud based or non-cloud-based is going to have an expected timeframe with failure-free operations. This assumption for reliability does get convoluted in a cloud computing environment considering the infrastructure is hidden to the subscriber. Regardless, there are four individual components for cloud reliability: (1) the hardware and software facilities offered by providers, (2) the provider's personnel, (3) connectivity to the subscribed services, and (4) the subscriber's personnel [3]. When considering risks toward the use of cloud computing, connectivity to the subscribed services, and the subscriber's personnel are going be a constant risk for the whole organization regardless if cloud computing is used, therefore that leaves only the risk associated with the hardware and software facilities offered by providers and the provider's personnel. Service-level agreements have to be established to provide a baseline for cloud provider outages. Understanding the frequency and duration of cloud technology provider outages allows the organization to plan and provide resilient alternatives for any prolonged outage that are due to man-made or natural causes [3].

## 7. ECONOMIC GOALS

Economies of scale cloud computing promotes small up-front costs to research, develop and produce any application or new idea including a video surveillance management system. However, cloud technologies do not make it completely economically risk-free. In any marketplace, businesses can go out of business and that can hold true in the cloud technology space as well. Therefore, it has to be established when using cloud computing resources that there can be a risk involved directly related to the business continuity if a cloud computing vendor shuts down. Aside from the cloud computing vendor going out of business scenario, there are the traditional issues such as redundancy, replication, and diversity to help protect against physical and electronic mishap [3]. Since most cloud technology provides the protection layer for physical mishap, extra costs could be associated to provide a disaster recovery environment, which could include using different cloud technology providers.

## 8. COMPLIANCE

The ultimate responsibility for compliance in a cloud computing environment is going to be the cloud computing subscriber, with the cloud computing provider being in the best position to enforce compliancy rules [3]. With cloud computing environments, there is a sense of location independence that hides the knowledge of the exact location of the provider resources, which produces this lack of visibility into how the cloud functions. Therefore, when applying a video surveillance system using cloud technologies, this lack of invisibility can bring up the risk of not knowing the physical data locations which may add complexity to comply to a variety of different International and Federal jurisdiction and regulations or state statutes and directives[3]. This is going to have a direct impact when video is needed as part of an incident respond effort. Extracting video footage from a video surveillance management system on a cloud computing platform could result in following multiple state and federal regulations. Therefore it would be imperative to have an organization policy that supports a superset of state regulations for tracking and handling video footage from a video surveillance management system using cloud technologies.

## 9. DATA AND APPLICATION SECURITY

Data and application security needs of the organization for a video surveillance system are to be architecture and to provide a security framework that protects the various endpoints within the sphere of use, which illustrates the different ways people can access the information [4]. Figure 1 illustrates the various endpoints and layers that are needed to protect and the various security perimeters that define the outer limits of an organization's security and the beginning of the external world [4].

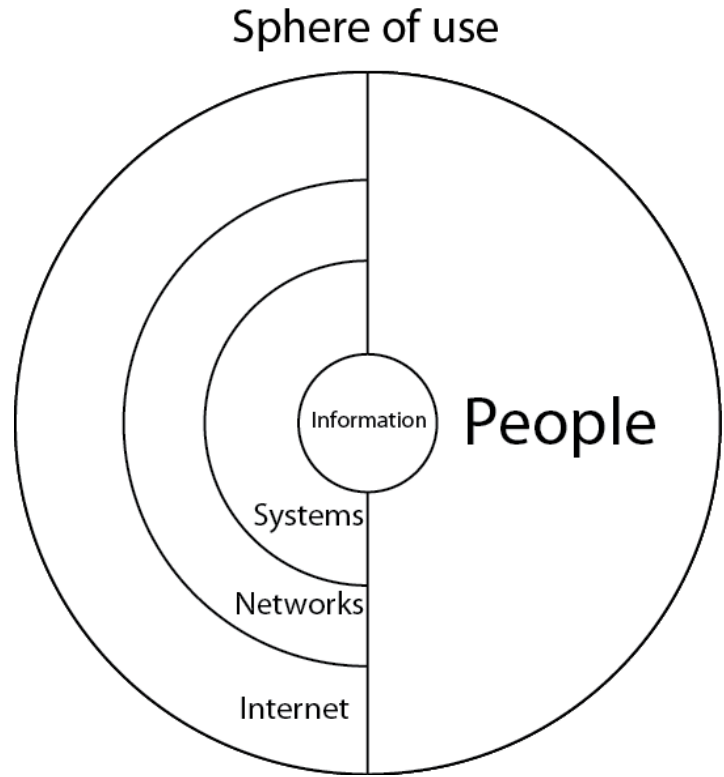

Figure 1. Sphere of use: Source[4]pp. 204.

The security framework will also define and provide the minimum levels of controls to provide a layered approached called defense in depth, to provide various levels of protection, and define the actual security perimeter around and outside the organization[4]. The NIST SP 800-14 provides an excellent outline for establishing acceptable practices for ensuring the security needs for the video surveillance system using cloud technologies. The security framework for the video surveillance system should start off with the following common security components for an establishing all security needs for external and internal use[5]:

- Policy—the policy program should set the organizational strategic direction to assign responsibilities and define a computer security program.
- Program Management—the program management functions are going to be a central security program. This will provide a long-term computer security strategy that is able to create external-organizational and intra-organizational liaison relationships.
- Risk Management—risk management is the process of assessing the organization's risk and take steps to reduce risk toward an acceptable level and maintaining that level of risk over the life of the company.

- Life Cycle Planning—is establishing best security practices during all the phases of an IT system life cycle: initiation, development/acquisition, implementation, operation, and disposal.
- Personnel/User Issues—no IT system can be secure without properly addressing staffing, separation of duties, least privilege, and user account management, unfriendly and friendly termination.
- Contingencies and Disasters—contingency planning addresses how to keep the organization's critical functions operating in the event of a disaster, both large and small. Mission-critical or business-critical functions should be identified and a BCP and DR plan should be created to maintain those systems up and running during disruption of services.
- Security Incident Handling— incident handling capabilities need to be defined on how to handle the incident when a computer virus, malicious code, or an outside or inside system intruder is detected.
- Awareness and Training—with proper planning, implementation, maintenance, and periodic evaluation, an effective computer security awareness and training program can be created that identifies the organization's security scope, goals, and objectives, that motivates management and employees to maintain a proper security framework.
- Security Considerations in Computer Support and Operations— a significant weakness is failing to consider security as part of the support and operations of IT systems. Poor documentation and no control of maintenance accounts can often lead to handicapping good security practices, therefore user support, software support, configuration management, backups, media controls, documentation, maintenance, and standardized log-on banners should all follow best practices for maintaining tight security architecture.
- Physical and Environment Security—physical and environmental security controls are implemented to protect the facility housing system resources, which can help prevent theft, unauthorized disclosure of information, interruptions in computer services, physical damage, loss of control over system integrity.
- Identification and Authentication—basis for most types of access controls and for establishing user accountability, is the critical building block of identification and authentication. Linking activities on an IT system to specific individuals establishes accountability, which requires the system to be able to identify users and differentiate between them to maintain a stable access control system.
- Logical Access Control—logical access controls should balance the often-competing interests of security, operational requirements, and user-friendliness in a way to explicitly enable or restrict user and system access to the organization's computer resources.
- Audit Trails— audit trails can provide a means to help accomplish several security-related objectives: reconstruction of events, individual accountability, intrusion detection, and problem identification by maintaining records of system activities, application processes or users.
- Cryptography—cryptography can be used to provide security services ensuring that data has not been modified and remains confidential. Cryptography methods should be implemented with standards that comply with export rules, manage keys and prevents the secure cryptographic Modules from being exploited.

By using cloud computing technologies for video surveillance system, eliminates any "physical separation" of user workloads and must rely on a "logical separation" to protect subscriber's resources [3]. Therefore, multi-tenancy has to be considered a security concern depending on the SPI model because there has been flaws documented using virtual machine technology [6].

## 10. SUMMARIZE TYPES OF EMERGING THREATS AND COUNTERMEASURES

Emerging threats that extend toward cloud computing environments are system complexity, shared multi-tenant environment, internet-facing services, and loss of control[7]. With cloud computing technologies, the number of interactions between components increases sufficiently which produces more probability to attack vectors, which could lead to rising chances to vulnerabilities[7]. Additionally, sharing resources with outside third parties on the same hardware requires cloud-computing providers to create logical separation within their infrastructure, which could lead to unauthorized access to all shared multi-tenants by exploitable vulnerabilities from within the cloud environments[7]. Previously defended intranet perimeters get less effective when that security layer is extended to the internet-facing services of cloud computing. Increased risk from network threats can rise due to moving existing services outside of the organization's protected intranet [7]. Finally, there are the additional security and privacy concerns when cloud technologies require a transfer or complete release of control over the organization's data. This loss of control can reduce the organization's ability to keep a valid situational awareness that involves their data being stored or processed with cloud-computing technologies [7].

NIST Special Publication 800-144 provides an excellent summary of recommendations to follow when there is an initiative to use cloud computing technologies, which is provided in Table 1 [7].

Table 1.Security and Privacy Issues and Recommendations Source [7]

| Areas | Recommendations |
|---|---|
| Governance | Extend organizational practices pertaining to the policies, procedures, and standards used for application development and service provisioning in the cloud, as well as the design, implementation, testing, use, and monitoring of deployed or engaged services. Put in place audit mechanisms and tools to ensure organizational practices are followed throughout the system lifecycle. |
| Compliance | Understand the various types of laws and regulations that impose security and privacy obligations on the organization and potentially impact cloud computing initiatives, particularly those involving data location, privacy and security controls, records management, and electronic discovery requirements. Review and assess the cloud provider's offerings with respect to the organizational requirements to be met and ensure that the contract terms adequately meet the requirements. Ensure that the cloud provider's electronic discovery capabilities and processes do not compromise the privacy or security of data and applications. |
| Trust | Ensure that service arrangements have sufficient means to allow visibility into the security and privacy controls and processes employed by the cloud provider, and their performance over time. Establish clear, exclusive ownership rights over data. Institute a risk management program that is flexible enough to adapt to the constantly evolving and shifting risk landscape for the lifecycle of the system. Continuously monitor the security state of the information system to support on-going risk management decisions. |
| Architecture | Understand the underlying technologies that the cloud provider uses to provision services, including the implications that the technical controls involved have on the security and privacy of the system, over |
| | the full system lifecycle and across all system components. |
| Identity and Access Management | Ensure that adequate safeguards are in place to secure authentication, authorization, and other identity and access management functions, and are suitable for the organization. |
| Software Isolation | Understand virtualization and other logical isolation techniques that the cloud provider employs in its multi-tenant software architecture, and assess the risks involved for the organization. |
| Data Protection | Evaluate the suitability of the cloud provider's data management solutions for the organizational data concerned and the ability to control access to data, to secure data while at rest, in transit, and in use, and to sanitize data. Take into consideration the risk of collating organizational data with that of other organizations whose threat profiles are high or whose data collectively represent significant concentrated value. Fully understand and weigh the risks involved in cryptographic key management with the facilities available in the cloud environment and the processes established by the cloud provider. |
| Availability | Understand the contract provisions and procedures for availability, data backup and recovery, and disaster recovery, and ensure that they meet the organization's continuity and contingency planning requirements. Ensure that during an intermediate or prolonged disruption or a serious disaster, critical operations can be immediately resumed, and that all operations can be eventually reinstituted in a timely and organized manner. |
| Incident Response | Understand the contract provisions and procedures for incident response and ensure that they meet the requirements of the organization. Ensure that the cloud provider has a transparent response process in place and sufficient mechanisms to share information during and after an incident. Ensure that the organization can respond to incidents in a coordinated fashion with the cloud provider in accordance with their respective roles and responsibilities for the computing environment. |

## 11. THE EVOLVING GIANT

There are two prospectives when implementing a video surveillance system, installing or upgrading. For the enterprise that is planning to install a video surveillance system, it is common to go directly to IP cameras and skip the whole closed-circuit video transmission (CCVT) and digital video recorders (DVR) eras. The biggest reasons why CCVT and DVR systems have a higher cost of installing cables to support both power and video is that they both lack the ability to scale upwards and lack interoperability [8]. Therefore, it is very common if not best practice today to install internet protocol (IP) cameras for a video surveillance system due to the advantages that power and video can be sent and received over Ethernet, which most enterprise company have installed already [8]. Combining IP cameras on the local network with a centralized video management software package and you have an enterprise video surveillance system that supports interoperability between vendors and is able to scale upwards for future growth. As good as an IP camera system sounds; there are two key resources that need to be examined with any IP camera system, bandwidth and storage [8]. According to one of the industry's leading IP-based megapixel camera manufacturers a 2 Megapixel camera can consume up to 3.1 Mbps of an enterprise's network [9]. For full details of the Arecont IP-camera, see

International Journal on Cryptography and Information Security (IJCIS),Vol.2, No.3, September 2012

Appendix A. Therefore, a building with ten 2 Megapixel cameras can consume up to 31 Mbps when using the H.264 codec, which on today's local Gigabit networks might not necessarily be a problem, but on networks with lower bandwidth capacity it could be a substantial problem to the point that it will not work with any success. Consequently, when using IP-cameras with high-resolution, using high bandwidth follows high storage requirements. Continuing to use the example of ten 2 Megapixel cameras, it would require 335GB of storage to store 24 hours of video footage at a 31 Mbps per camera. Therefore, it is easy to illustrate that a company that has over 1000 cameras will need over 3.35TB of storage to hold all the footage from the video surveillance system within a 24-hour period. Therefore it is important to allocate storage requirements for a video surveillance system based on actually factoring in the following parameters [10]:

- Number of cameras
- Whether recording will be continuous or event-based
- Number of hours per day the camera will be recording
- Frames per second
- Image resolution
- Video compression type: Motion JPEG, MPEG-4, H.264
- Scenery: Image complexity (e.g. gray wall or a forest), lighting conditions and amount of motion (office environment or crowded train stations)
- How long data must be stored

From the list above, motion detect is the parameter that is variable by nature. All the other parameters get defined by policy, standards and guidelines, which places them as static values. Normally, when installing or upgrading a video surveillance system, configuration such as, frames per second, image resolution, video compression, storage retention, and number of camera is determined and becomes the standard across the whole enterprise. Therefore, to help with this huge burden with storage, most video surveillance system provides motion detection, which only stores video footage when it detects motion, thus it is a better use of storage for video footage that is based on movement rather than to waste storage on footage that shows the same hallway for 24-hours, as this can significantly reduce that amount of storage needed. Of course, the trick here is the ability to accurately predict the percentage of motion per camera. With that information along with all the other parameters, it would be a lot easier to estimate storage requirements for the video surveillance system.

**11.1 Research Parameters for Cloud Technologies**

For the purpose of this research, the parameters for the current video surveillance system, which can affect storage and bandwidth, will be set to the following:

- There will be 278 cameras
- Recording will be continuous using motion detection
- Will be recording 24-hours a day
- Frames per second will be set to 15 fps
- Image resolution will be set to 1600 pixels X 1200 pixels
- Video compression type used will be set to H.264 medium quality
- Data must be stored for 14 days.

It is important to note that the setting for frames per second is a significant setting, considering that is the baseline to be used to determine the quality and bandwidth consumption for an IP-camera. Using the Arecont AV2105 as the standard IP-camera in an enterprise it is possible to have a 1600 pixel X 1200 pixel video at 24 fps; however, it will consume more power and bandwidth to maintain such a higher quality video image. Therefore, 1600 pixel X 1200 pixel video at 15 fps is a good compromise between video quality and camera resources (Arecont Vision, 2012). Additionally, there is also a huge advantage to using the H.264 codec to provide the video stream to a video management system. To record video at a 1600x200 resolution will require the following bandwidth based on raw footage:

1. 1600 pixels X 1200 pixels = 1,920,000 pixels per Image
2. 1,920,000 pixels per Image X 24 bits/pixel = 46080000 bits per Image
3. 46080000 bits per Image / 1000 / 1000 = 46.08 Mb per Image
4. 46.08 Mb per Image X 15 Images per sec = 691.2 Mbps

Therefore, it would require 691.2 Mbps to stream the raw video data with a resolution at 1600x1200 pixels. According to Arecont by using H.264 (MPEG4 Part 10) Compression Standard on a medium setting for quality can reduces the raw data size per image from 46.08 Mb to 0.0267 Mb per image, which brings down the bandwidth usage to 3.1 Mbps when using H.264 [9]. Table 2 highlights resources usage from the current video surveillance system that is collecting video footage globally in eight different countries for a specific enterprise.

Table 2: Current system usage baseline for resources when comparing cloud technologies between locally deployed video management systems.

| Server Name | Cameras | Storage Used (GB) | Viewing 25% of Cameras Bandwidth (Mbps) | Storage per Day (GB) | Camera Bandwidth (Mbps) | Calculated % Motion | Average Camera Footage per Day (GB) |
|---|---|---|---|---|---|---|---|
| RecSvr1 | 92 | 2930 | 71.3 | 209.29 | 285.2 | 44.62 | 2.27 |
| RecSvr2 | 8 | 1974 | 6.2 | 141.04 | 24.8 | 30.07 | 17.63 |
| RecSvr3 | 73 | 1880 | 58.9 | 134.29 | 226.3 | 28.63 | 1.84 |
| RecSvr4 | 25 | 156 | 21.7 | 27.53 | 77.5 | 2.38 | 0.45 |
| RecSvr5 | 8 | 385 | 6.2 | 27.53 | 24.8 | 5.87 | 3.44 |
| RecSvr6 | 5 | 31 | 6.2 | 2.27 | 15.5 | .48 | 0.45 |
| RecSvr7 | 9 | 79 | 9.3 | 5.71 | 27.9 | 1.22 | 0.63 |
| RecSvr8 | 12 | 77 | 9.3 | 5.51 | 37.2 | 1.17 | 0.46 |
| RecSvr9 | 26 | 637 | 21.7 | 45.59 | 80.6 | 9.70 | 1.75 |
| RecSvr10 | 6 | 333 | 6.2 | 23.84 | 18.6 | 5.08 | 3.97 |
| RecSvr11 | 14 | 767 | 12.4 | 54.79 | 43.4 | 11.68 | 3.91 |
| Averages | | | | 61.58 | | 12.8 | 3.345 |
| Totals | 278 | 9249 | 229.4 | | 861.8 | | |

Table 3 highlights the current resources costs for implementing an enterprise video surveillance system that is able to collect video footage globally in eight different countries.

Table 3. Current System Cost for Comparing with Cloud Technologies

| | Description | Qty. | Cost | Total |
|---|---|---|---|---|
| **Hardware** | Recording Servers | 11 | $17,000.00 | $187,000.00 |
| **Software** | Licenses | 278 | $310.00 | $86,180.00 |
| | Application | 1 | $5,000.00 | $5,000.00 |
| **Power*** | Hardware & Cooling | 11 | $806.96 | $8,876356 |
| **First Year Total** | | | | $278,180.00 |
| **Yearly Expense** | | | | $95,056.56 |

**11.2 CLOUD SERVICE MODELS**

Cloud computing technologies is the evolutional direction for the computing world to provide network accessible, convenient, ubiquitous, on-demand configurable computing resources that can be provisioned rapidly with minimal efforts by management and service [2]. The National Institute of Standards and Technology (NIST) have composed five essential characteristics for cloud computing [2]:

1. On-demand self-service: Unilaterally provision automatically without requiring human interaction.
2. Broad network access: Capabilities are available over the network.
3. Resource pooling: User generally is not aware of the exact location of the provided resources creating a sense of location independence.
4. Rapid elasticity: Resources can be elastically provisioned and released as the demand for resources increase or decrease.
5. Measured service: Resources can be monitored, controlled, and reported creating a transparency level of abstraction for metering capabilities consumed for both the users and providers.

Cloud computing model spans across three different types of service models for delivering the five essential characteristics of cloud computing, Software as a Service (SaaS), Platform as a Service (PaaS), Infrastructure as a Service (IaaS) [2]. Software as a Service provides applications running on a cloud infrastructure that can be accessible by various client devices [2]. With a SaaS service model, the underlying cloud infrastructure is completely out of reach to the users, they only have access to the application on the cloud infrastructure. SaaS could be considered the equivalent to Commercial Off-the-Shelf (COTS) software but running on a cloud-computing model. Now to take the SaaS service model to include highly customizable software would lead toward the Platform as a Service (PaaS) service model. The PaaS service model provides a platform to create custom applications from the ground up using the PaaS provided programming languages, services, libraries and tools that is specific to that PaaS provider. With a PaaS service model, the underlying cloud infrastructure is completely out of reach to the users; however, the PaaS provider will allow the users the ability to control how their application's deployment over their cloud infrastructure [2]. If more control of the application is required, such as provisioning processing, storage, and networking resources, then an Infrastructure as a Service (IaaS) service model provides the consumer additional options. With a IaaS service model, the underlying cloud infrastructure is completely out of reach to the users; however, the IaaS provide will allow the users the ability to granularly deploy operating systems, networking components, storage, and deployable applications to create cloud-computing solutions [2]. Figure 2 outlines the various controls from the consumer and providers point of view across the different SaaS, PaaS, and IaaS (SPI) models.

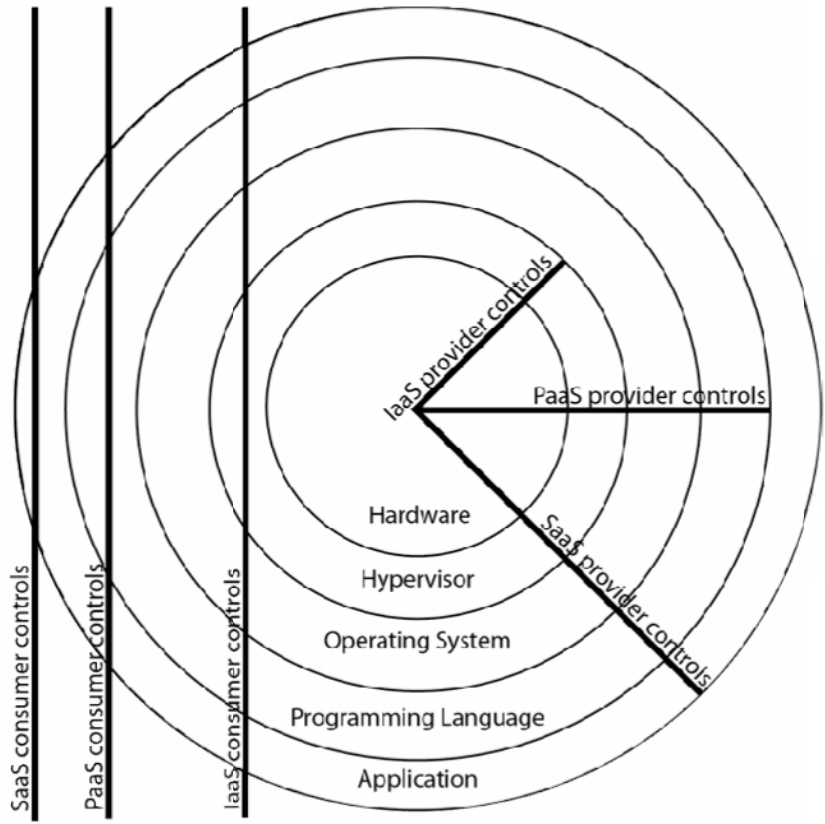

Figure 2. SPI Model Access Controls.

**11.2.1 Software as a Service**

To use Software as a Service (SaaS) solution for the configuration above would result in changing the requirements. Most Video Surveillance as a Service (VSaaS) providers uses Axis Video Hosting System and is configured by various security integrators and alarm monitoring centers to provide services [10]. The requirement to use 2MP cameras will not be fulfilled by using a SaaS solution; in fact, by using a Video Surveillance as a Service (VSaaS) system, choices are going to be limited to hardware that is compatible to the VSaaS systems. For the most part the VSaaS providers are compatible to certain cameras supporting 640x480-resolution video at 5-6 fps using H.264 codex for streaming video to their services [10]-[11]. To use a SaaS service would require purchasing new cameras or if possible setup and configure all 278 cameras to be compatible with the VSaaS provider and include their monthly fee per camera which is estimated at $30.77 for a month to store 14 days of video [11]. A monthly bill estimated at $8554.06 a month would be the cost, which would create a total yearly cost at $102,648.72 to use a SaaS service. By using a VSaaS provider for a VMS provides the only added security benefits: provides off-premise video storage to reduce the chance of theft and storing video footage in a SAS 70, RSA Encryption or ISO 27001-compliance infrastructure. However, there are additional security risks by using a VSaaS; reduced defense layer by having only having one layer of access control to the camera that is directly exposed to the internet, lack of camera support for encrypting the video feed using IPSec/VPN or SSL back to the VSaaS providers to prevent ease-dropping.

International Journal on Cryptography and Information Security (IJCIS),Vol.2, No.3, September 2012

**11.2.2 Platform as a Service**

To use a Platform as a Service (PaaS) solution for the configuration above would result in more planning toward choices in configuration and application programming languages to meet the requirements. By using a PaaS solution, the goal is to create a custom application to solve the organization's specific problem or create a SaaS environment to leverage the solutions to multiple organizations. The PaaS providers that have the strongest market share are Microsoft Windows Azure, SalesForce.com, and Google App Engine. PaaS providers like OrangeScape and Wolf Framework are adding another layer in the PaaS cloud by focusing more on preventing vendor lock-in by allowing deployment with any of the major cloud platforms or deploying the same application without any change in private cloud infrastructure or data center using their products [12]-[13]. Using the pricing model for Windows Azure PaaS with the sample of 248 cameras. Table 4 outlines the services and costs[14].

Table 4. Windows Azure Estimated Price Breakdown for a PaaS VMS (Pricing was taken on February 10th 2012 from http://www.windowsazure.com/en-us)

| Service | Description | Qty. | Cost | Total |
|---|---|---|---|---|
| Compute | | | | |
| | Small Computer | 5 | $90.00 | $450.00 |
| | Medium Computer | 2 | $180.00 | $360.00 |
| | Large Computer | 2 | $360.00 | $720.00 |
| | Extra Large Computer | 2 | $720.00 | $1,440.00 |
| SQL Azure Database | 20 GB | 1 | $65.94 | $65.94 |
| Blob, Table and Queue Storage | | | | |
| | Storage 9450 GB | 1 | $1,323.01 | $1,323.01 |
| | 6 Million Storage Trans | 1 | 1 | $1.00 |
| Bandwidth | Based 25% viewing of video | | | |
| | 1000 GB North America + Europe Egress | 50 | $120.00 | $6,000.00 |
| | 1000 GB Other Locations Egress | 20 | $190.00 | $3,800.00 |
| Service Bus | 1 Million Messages | 2 | $2.00 | $4.00 |
| Access Control | 1 Million Access Control Transactions | 1 | $19.90 | $19.90 |
| Total per month | | | | $14,183.85 |
| Yearly Total | | | | $170,206.20 |

Only cost of resource consumption used with a PaaS solution using Windows Azure is indicated in Table 2 and it does not reflect the cost of resources for creating the actual application using the PaaS cloud infrastructure. The assumption for such a task to produce a VMS on a PaaS cloud infrastructure could be introduced to the current vendor leaders in the industry. However to simplify, that cost would have to be added to the total cost for a PaaS application solution.

**11.2.3 Infrastructure as a Service**

To use an Infrastructure as a Service (IaaS) solution for the configuration above would result in more planning toward choices in configuration and application programming languages to meet the requirements. Some of the top Infrastructure as a Service (IaaS) providers are Amazon's Elastic Compute Cloud (EC2), GoGrid and Rackspace, Terremark[15]. Using the pricing model for Amazon's Web Services for an IaaS solution for the sample of 248 cameras.

International Journal on Cryptography and Information Security (IJCIS),Vol.2, No.3, September 2012Table 5 outlines the services and costs [16].

Table 5. Amazon Elastic Compute Cloud Estimated Price Breakdown for a IaaS VMS (Pricing was taken on February 18th 2012 from http://aws.amazon.com)

| Service | Description | Qty. | Cost | Total |
|---|---|---|---|---|
| Compute | | | | |
| | Standard On-Demand Instances Extra Large | 5 | $0.96 per hour | $3,456.00 |
| | Hi-Memory On-Demand Instances Extra Large | 2 | $0.62 per hour | $892.80 |
| | Hi-Memory On-Demand Instances Double Extra Large | 2 | $1.24 per hour | $1,785.60 |
| | Hi-Memory On-Demand Instances Quadruple Extra Large | 2 | $2.48 per hour | $3,571.20 |
| Amazon Relational Database Service (Amazon RDS) | | | | |
| | Standard Deployment Small DB Instance | 1 | $0.11 per hour | $79.20 |
| | Standard Deployment Storage Rate | 20 | 0.10 per GB-month | $2.00 |
| Amazon Simple Storage Service (Amazon S3) | | | | |
| | Over 5000 TB / month | 9450 | $0.055 per GB | $519.75 |
| | 1,000 request | 6,000 | $0.01 | $60.00 |
| Bandwidth | Based 25% viewing of video | | | |
| | Next 350 TB / month | 70000 | $0.050 per GB | $3,500.00 |
| Amazon Simple Queue Service (Amazon SQS) | First 1 GB / month | 1 | $0.000 per GB | $0.00 |
| AWS Identity and Access Management (IAM) | Free | 1 | $0 | $0.00 |
| Total per month | | | | $13,866.55 |
| Yearly Total | | | | $166,398.60 |

Only cost of resource consumption used with a IaaS solution using Amazon Web Services is indicated in Table 4 and it does not reflect the cost of resources for creating the actual application using the IaaS with a PaaS cloud infrastructure.

**11.2.3 CLOUD DEPLOYMENT MODELS**

Regardless of which SaaS, PaaS, IaaS service model (SPI) used, the cloud-computing infrastructure still needs deploying into solid ground regardless if it is labeled "in the cloud."

NIST identifies four deployment models for cloud computing; private cloud, community cloud, public cloud, and hybrid cloud [2]. NIST Special Publication 800-146 outlines the four deployment models as follows [3]:

- Private cloud. The cloud infrastructure is operated solely for an organization. It may be managed by the organization or a third party and may exist on premise or off premise.
- Community cloud. The cloud infrastructure is shared by several organizations and supports a specific community that has shared concerns (e.g., mission, security requirements, policy, and compliance considerations). It may be managed by the organizations or a third party and may exist on premise or off premise.
- Public cloud. The cloud infrastructure is made available to the general public or a large industry group and is owned by an organization selling cloud services.
- Hybrid cloud. The cloud infrastructure is a composition of two or more clouds (private, community, or public) that remain unique entities but are bound together by standardized or proprietary technology that enables data and application portability (e.g., cloud bursting for load-balancing between clouds).

The industry is rapidly growing with various deployment options that can support the full range of cloud computing services. There is going to be six options to choose from when deciding on deployment model architecture [3]:

1. On-Site Private Cloud
2. Outsourced Private Cloud
3. On-Site Community Cloud
4. Outsourced Community Cloud
5. Public Cloud
6. Hybrid Cloud: a mix of any of the five above.

## 12. ANALYZE CURRENT REAL WORLD TRENDS

Based on today's cloud computing resources that are available, it is possible to create a video management system in the cloud. It is not necessarily going to be the least expensive solution, based on the current pricing options that are available from the various cloud technology providers. With the extreme requirements toward video storage, the massive amounts of network usage when using IP-cameras is going to stress the limitations of any cloud technology architect. Table 6 breaks down a hybrid cloud computing solution for optimizing the least expensive solution; however, it takes advantage of using two different providers. Therefore, the hybrid solution provided by different vendors at this time would be considered as theoretical. Assuming the latency between using two separate providers would not necessarily produce a functional system for the purposes of a video surveillance system. Regardless, as more competitive prices are created between providers for less expensive cloud computing technologies, it starts to create a more viable solution that can be achievable.

Table 6.Estimated Price Breakdown for a combing a PaaS and IaaS solution for a VMS (Pricing was taken on February 10th 2012 from http://www.windowsazure.com/en-us/ and pricing was taken on February 18th 2012 from http://aws.amazon.com)

| Service | Description | Qty. | Cost | Total |
|---|---|---|---|---|
| Compute | | | | |
| | Small Computer | 5 | $90.00 | $450.00 |
| | Medium Computer | 2 | $180.00 | $360.00 |
| | Large Computer | 2 | $360.00 | $720.00 |
| | Extra Large Computer | 2 | $720.00 | $1,440.00 |
| SQL Azure Database | 20 GB | 1 | $65.94 | $65.94 |
| Amazon Simple Storage Service (Amazon S3) | | | | |
| | Over 5000 TB / month | 9450 | $0.06 | $519.75 |
| | 1,000 request | 6,000 | $0.01 | $60.00 |
| Bandwidth | Based 25% viewing of video | | | |
| | Next 350 TB / month | 70000 | $0.05 | $3,500.00 |
| Amazon Simple Queue Service (Amazon SQS) | First 1 GB / month | 1 | $0.00 | $0.00 |
| AWS Identity and Access Management (IAM) | Free | 1 | $0 | $0.00 |
| Total per month | | | | $7,115.69 |
| Yearly Total | | | | $85,388.28 |

When comparing the various SPI model architectures that are available based on today's market, it is clear that Table 7 illustrates the standard internal deployment that integrates into the current enterprise infrastructure still remains the least expensive solution based on the current requirements. The SaaS solution does provide a viable option, however it does have shortcomings that need to address the current requirements for using high-resolution IP-cameras. The PaaS and IaaS solutions currently are close to being twice as much in costs, which provides no savings over the long haul when compared to purchasing the hardware separately for a video surveillance system. Contrary, when a hybrid solution is architecture, using the various vendors to produce the lowest costs, there is an actual savings of $9668.28 a year. If it is possible to create a security video surveillance system for under $187,000.00, which is the purchase price for the hardware, then that would make a viable solution to architect using cloud technologies. One day in the future it might be possible to create a cloud computing solution for a video surveillance system that is less expensive than purchasing the actual hardware, however based on today's market a cloud computing solution is not the less expensive solution compared to purchasing the hardware and supplying power.

Table 7. Estimated Price Breakdown for a combined a PaaS and IaaS solution for a VMS

| Model | Provider | Monthly Cost | Yearly Cost |
|---|---|---|---|
| Standard | Standard Internal | $7,921.88 | $95,056.56 |
| SaaS | ipConfigure | $8,554.06 | $102,648.72 |
| PaaS | Windows Azure | $14,183.85 | $170,206.20 |
| IaaS | Amazon EC2 | $13,866.55 | $166,398.60 |
| Hybrid | Amazon EC2 & Windows Azure | $7,115.69 | $85,388.28 |

## 13. CONCLUSION

To deploy a video surveillance system into the cloud is possible but would inherent either the limitation of not being a high-resolution video surveillance system or the limitation based on availabilities due to the extreme resource requirements on video storage and network usage. Additionally, with the current pricing model available today and with the extreme amounts of storage and network bandwidth requirements for a high resolution IP-camera video surveillance system the cost for deploying or creating a video surveillance management system using cloud computing it is not going to be less expensive than purchasing and deploying the hardware locally.

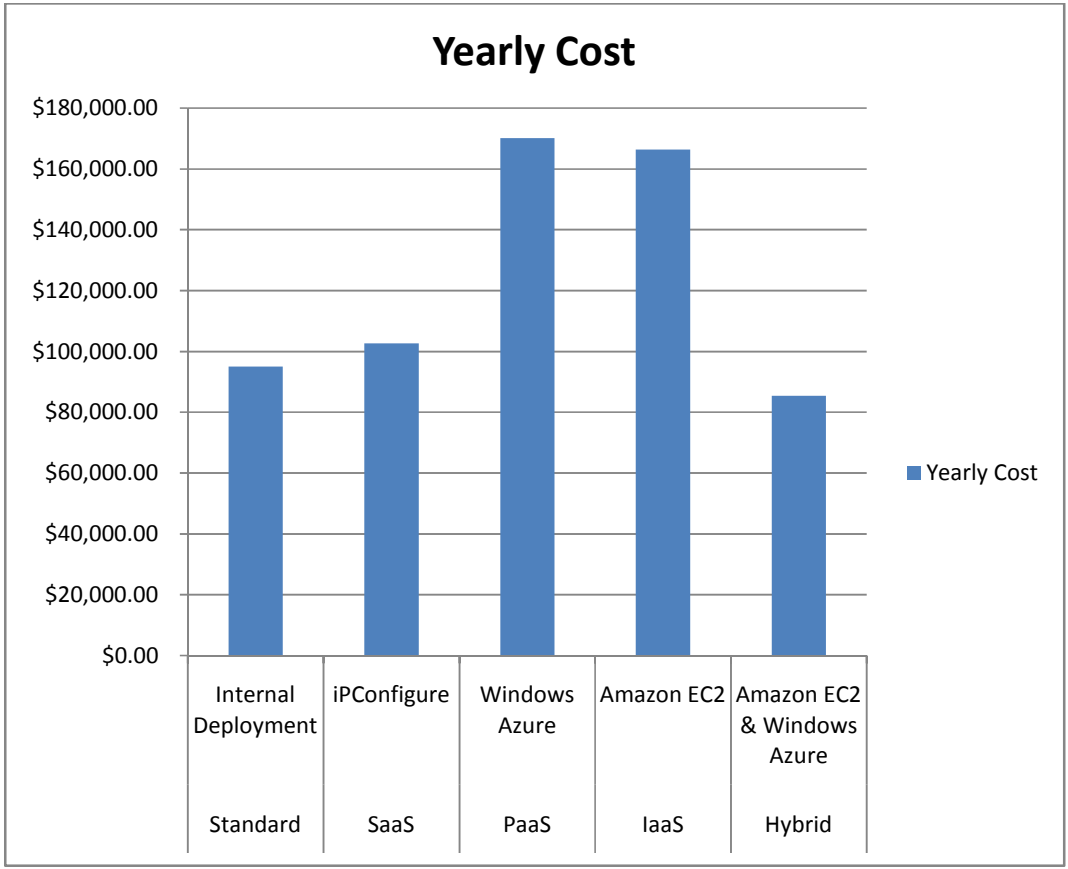

Figure *3*. Overview of Yearly Cost compared to all SPI Models

The security challenges that are presented in any enterprise environment are going to be extended to any cloud computing environment as well. Inherently, with the added feature of resource pooling that can present resource with no sense of location dependency will require the organization's security posture to extend to include the client and server sides of the cloud computing environments. Additionally, Figure 3 graphically illustrates just how much more expensive it will be to deploy a high-resolution video surveillance system using today's cloud-computing technologies compared to the locally deployed solutions, plus it illustrates how cost effective it can be by combining theoretically two separate cloud computing vendor solutions together to take advantage of available pricing.

## Authors' Bio

DJ Neal (Security+, Network+) holds a Master of Science Degree in Information Assurance and Security from Capella University (2012), Bachelor of Science in Computer Science from University of Las Vegas (2000), and an Associate Degree in Nuclear Technology from University of Phoenix (1997). DJ Neal's current interests include, networking, database security, cloud computing, security architecture, physical access controls, surveillance systems, and computer forensics.

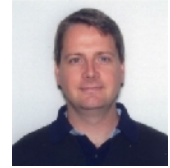

Syed (Shawon) M. Rahman is an assistant professor in the Department of Computer Science and Engineering at the University of Hawaii-Hilo and an adjunct faculty of information Technology, information assurance and security at the Capella University. Dr.Rahman's research interests include software engineering education, data visualization, information assurance and security, web accessibility, and software testing and quality assurance. He has published more than 75 peer-reviewed papers. He is a member of many professional organizations including ACM, ASEE, ASQ, IEEE, and UPE.

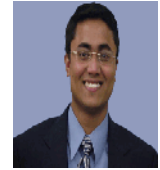